\documentclass[structabstract]{aa}  
\usepackage{graphicx}
\usepackage{rotating}
\usepackage[]{natbib}
\usepackage{txfonts}

\begin{document}
   \title{{\it XMM-Newton} and {\it Chandra} observations of G272.2-3.2. Evidence of stellar ejecta in the central region.
   }

   \author{E. S\'anchez-Ayaso\inst{1}, J.A. Combi\inst{2,6},  F. Bocchino\inst{3}, J.F. Albacete-Colombo\inst{4}, J. L\'opez-Santiago\inst{5}, J. Mart\'{\i}\inst{1}, and E. Castro\inst{5}
         }

\authorrunning{S\'anchez-Ayaso et al.
}

\titlerunning{{\it XMM-Newton} and {\it Chandra} observations of G272.2$-$3.2
} 

\offprints{S\'anchez-Ayaso}  

\institute{Deptamento de F\'isica (EPS), Universidad de Ja\'en, Campus Las Lagunillas s/n Ed. A3 Ja\'en, Spain, 23071\\
\email{[esayaso:jmarti]@ujaen.es}
\and
Instituto Argentino de Radioastronom\'{\i}a (CCT La Plata, CONICET), C.C.5, (1894) Villa Elisa, Buenos Aires, Argentina.\\
\email{jcombi@fcaglp.unlp.edu.ar}
\and
INAF-Osservatorio Astronomico di Palermo, Piazza del Parlamento 1, 90134, Palermo, Italy.\\
\email{bocchino@astropa.inaf.it}
\and
Centro Universitario Regional Zona Atl\'antica (CURZA). Universidad Nacional del COMAHUE, Monse\~nor Esandi y Ayacucho (8500), 
Viedma (Rio Negro), Argentina.\\
\email{donfaca@gmail.com}
\and
Departamento de Astrof\'{\i}sica y Ciencias de la Atm\'osfera, Universidad Complutense de Madrid, E-28040, Madrid, Spain.\\
\email{[jls:eli]@astrax.fis.ucm.es}     
\and         
Facultad de Ciencias Astron\'omicas y Geof\'{\i}sicas, Universidad Nacional de La Plata, Paseo del Bosque, B1900FWA La Plata, Argentina.
             }

   \date{Received 30 May 2012 ; accepted 4 January 2013 }

 
  \abstract
	{}
	{We aim to study the spatial distribution of the physical and chemical properties of the X-ray emitting plasma of the supernova remnant G272.2$-$3.2, in order to obtain important constraints on its ionization stage, the progenitor supernova explosion, and the age of the remnant.
	} 
   {We report on combined XMM-Newton and Chandra images, median photon energy maps, silicon and sulfur equivalent width maps, and a spatially resolved spectral analysis for a set of regions of the remnant. Complementary radio and H$\alpha$ observations, available in the literature, are also used to study the multi-wavelength connection of all detected emissions.
   }
	{
	The X-ray morphology of the remnant displays an overall structure with an almost circular appearance, a centrally brightened hard region, with a peculiar elongated hard structure oriented along the northwest-southeast direction of the central part. The X-ray spectral study of the regions shows distinct K$\alpha$ emission-line features of metal elements, confirming the thermal origin of the emission. The X-ray spectra are well represented by an absorbed variable abundance non-equilibrium ionization (VNEI) thermal plasma model, which produces elevated abundances of Si, S, and Fe in the circular central region, typical of ejecta material. The values of abundances found in the central region of the supernova remnant (SNR) favor a Type Ia progenitor for this remnant. The outer region shows abundances below the solar value, to be expected if the emission arises from the shocked interstellar medium (ISM). The relatively low ionization timescales suggest non-equilibrium ionization. We identify the location of the contact discontinuity. Its distance to the outer shock is higher than expected for expansion in a uniform media, which suggests that the remnant spent most of its time in more dense medium.
	}
{}

	\maketitle

\section{Introduction}

Supernova remnants (SNRs) are usually classified as a shell-like,  plerionic, or composite class, according to their radio and X-ray morphology. Although  the detection of non-thermal radio emission is considered a prerequisite for a true SNR identification (Green 2009), there are a number of SNRs that present weak or undetectable radio emission. These objects may belong to the so-called class of radio quiet SNRs (Mavromatakis \& Strom 2002, Mavromatakis et al. 2005, Boumis et al. 2002).   

The southern Galactic SNR G272.2$-$3.2 seems to be a member of this class. It was first detected at X-ray energies by Greiner \& Egger (1993) as part of the ROSAT All-Sky Survey and later studied in more detail by Greiner, Egger \& Aschenbach (1994). These authors observed that it has an almost circular X-ray morphology with a diameter of about 15 arcmin. A more recent X-ray study of G272.2$-$3.2 was carried out by Harrus et al. (2001), using a combination of ROSAT and ASCA data. They found that this object has a centrally brightened morphology and thermally dominated X-ray emission and therefore classified the object as a ``thermal composite" SNR, a class of SNR still poorly understood. 

\begin{table*}
\caption{{\it Chandra} and {\it XMM-Newton} observations of G272.2$-$3.2.}
\label{obs}\centering
\begin{center}
\begin{tabular}{l c c l c c c c}
\hline\hline
Satellite& \multicolumn{2}{c}{{\it Chandra}}&& \multicolumn{1}{c}{{\it XMM-Newton}} \\ 
\cline{2-3} \cline{5-6}
Obs-Id                   & 9147 & 10572 && 0112930101 \\ 
Date	                & 27/08/2008  & 28/08/2008	&& 10/12/2001 \\
Start Time [UTC]   & 08:12:00   & 23:39:00 && 23:01:35	\\
Camera    	         & ACIS-012367 & ACIS-012367	&& MOS1,2/pn		\\
Filter 		         &  $--$	       	&	$--$			&& Thin	\\
Modes (read/data) & TIMED/VFAINT & TIMED/VFAINT && PFWE			\\
Offset 		         &   on axis   & on axis && on-axis	\\
Exposure [ks]        & 29.71   & 	33.66	&& 36.8-37.3/27.9   	\\
GTI	[ks]	               & 26.04   & 30.59      && 29.5-30.2/24.4   &	\\
\hline
\end{tabular}
\end{center}
\tablefoot{All observations were taken from the respective mission archives. PFWE refers to the Prime Full Window Extended observation mode. The pointing of {\it Chandra} observations is centered at the following J2000.0 coordinates: $\alpha$= 09$^h$06$^m$47$\fs$00, $\delta$=-52$\degr$05$\arcmin$50$\farcs$0.}
\label{obstable}
\end{table*}

\begin{figure*}
\centering
\includegraphics[width=9cm]{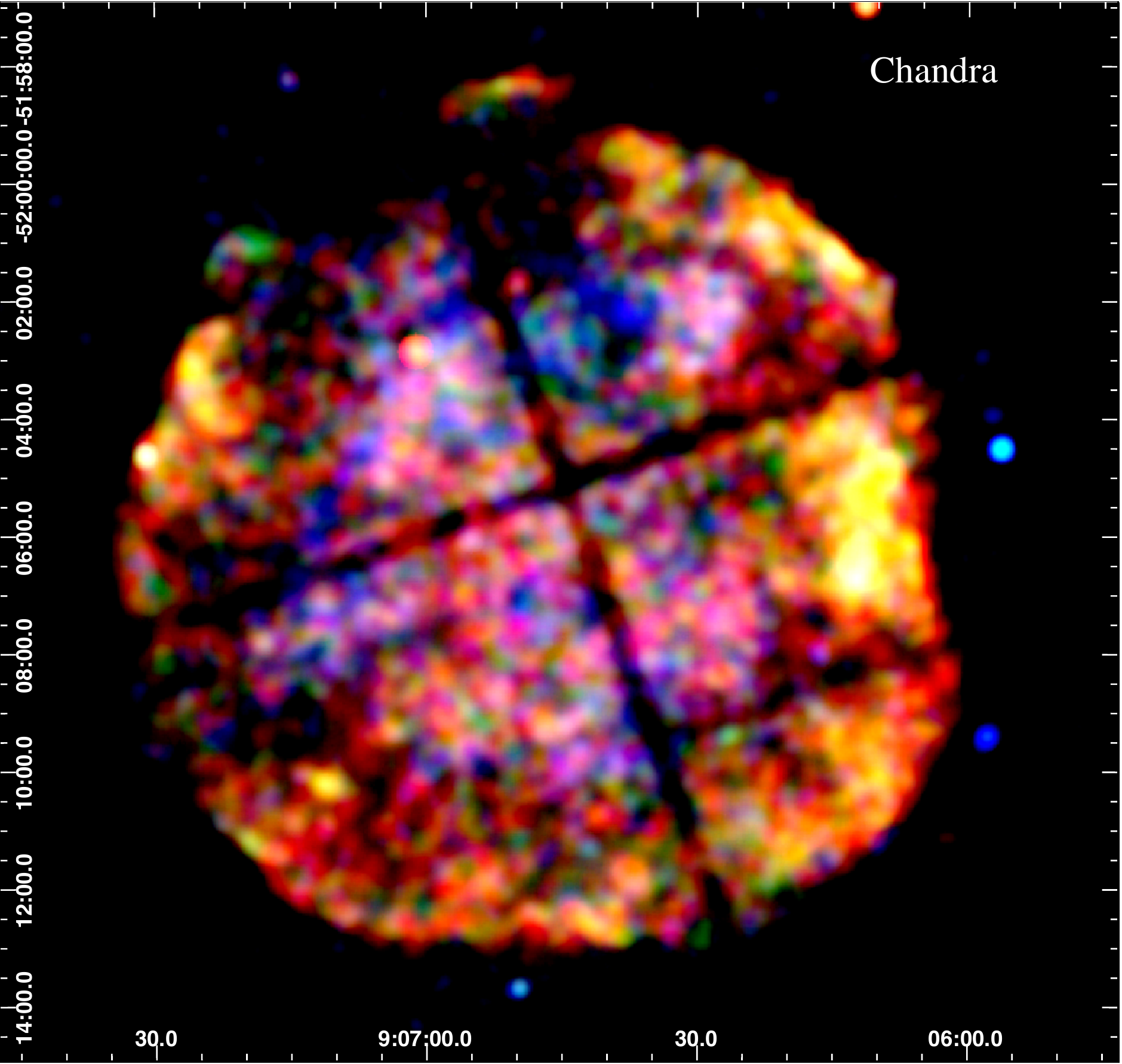}
\includegraphics[width=9cm]{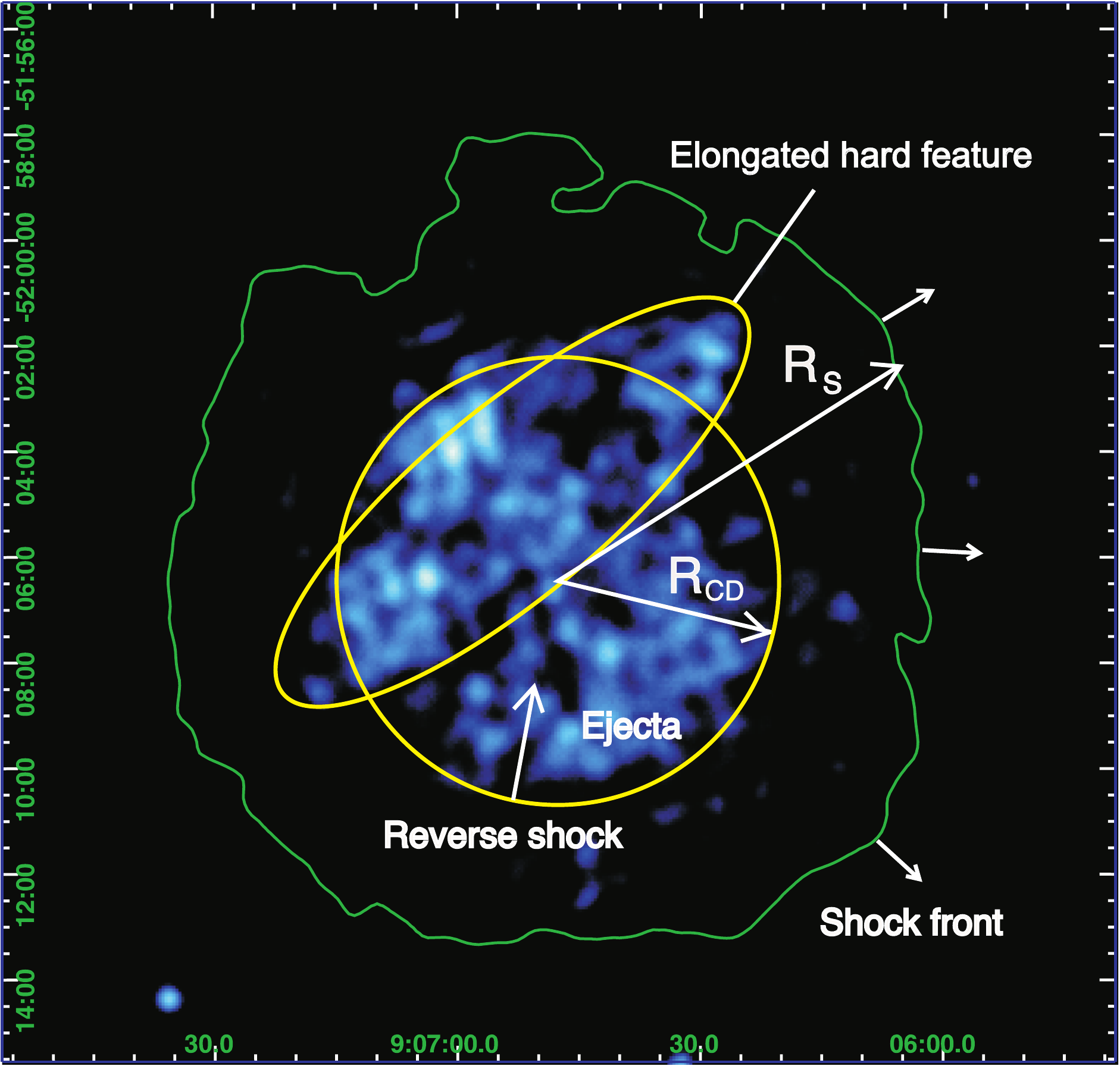}
\caption{{\bf Left panel:} {\it Chandra} image, covering a 16$\times$16 arcmin$^{2}$ field of view, of G272.2$-$3.2 in the three X-ray energy bands: soft (0.5-1.0 keV) in red, medium (1.0-1.8 keV) in green, and hard (1.8-3.0 keV) in blue. The image corresponds to the merged {\it Chandra} observations. {\bf Right panel:} Combined {\it XMM-Newton} MOS1 and MOS2 image in the hard X-ray energy range (1.8-3.0 keV). The boundary of the SNR is indicated in yellow. The circle in the central part of the SNR indicates the reverse shock interface.  The region of the elongated hard emission is also indicated by a yellow ellipse. $R_{\rm CD}$ and $R_{\rm S}$ indicate the radii of the contact discontinuity and the forward shock front, respectively. Horizontal and vertical axes are labeled using J2000.0 right ascension and declination.}
\label{Fig2}
\end{figure*}

Greiner, Egger \& Aschenbach (1994) also analyzed optical  observations of G272.2$-$3.2 from the ESO/SRC plate R 6712. They found that there exists a faint nebulosity near the center and an extended component on the west part of the SNR. At radio frequencies, the SNR was first studied by Duncan et al. (1997), using the Parkes, ATCA, and MOST radiotelescopes. In radio, the object displays a low surface brightness (there is no clear evidence of a shell-like morphology) with a steep non-thermal radio spectral index of $-$ 0.55$\pm$0.15 (S$\propto$ $\nu^{\alpha}$). Evidence of polarized emission or a pulsar wind nebula was not found in this study.

Recently, Sezer \& G$\ddot{\rm o}$k (2012) presented Suzaku observations of G272.2$-$3.2. The authors found that the X-ray spectrum is well fitted by a single-temperature variable abundance non-equilibrium ionization (VNEI) model with an electron temperature of kT $\sim$ 0.77 keV. Enhanced abundances of Si, S, Ca, Fe, and Ni were observed in the central region, indicating that the X-ray emission has an ejecta origin. The relative abundances found in the central region suggest that G272.2$-$3.2 is the result of a Type Ia supernova explosion.

The distance to G272.2$-$3.2 is uncertain. Analyzing the interstellar absorption, Greiner, Egger \& Aschenbach (1994) obtained a distance of 1.8$^{+1.4} _{-0.8}$ kpc. Using statistical analysis, Harrus et al. (2001) located the SNR at 2 kpc, with an upper limit of 10 kpc. They adopted an intermediate distance of 5 kpc. Taking this distance into account, Koo et al. (2004) found an excess of HI emission (see their Fig. 2), which peaks at the western part of the remnant, where an infrared shell was detected by Harrus et al. (2001). This result suggests that G272.2$-$3.2 interacts with dense interstellar medium (ISM). 

In this paper, we report the results of a combined analysis of {\it XMM-Newton} and {\it Chandra} observations of G272.2$-$3.2. The paper is organized as follows: in Sect.~2, we describe the {\it XMM-Newton} and {\it Chandra} observations and the data reduction process. In Sect.~3, we present the main results of our X-ray data analysis, including X-ray images, spectra, and a mean photon energy map. In Sect. 4, we discuss a possible scenario to explain the nature of the observed characteristics of the remnant, and in Sect.~5, we summarize our conclusions.

\section{X-ray observations and data reduction}

We combined {\it XMM-Newton} and {\it Chandra} data in order to carry out a detailed spectral and spatial X-ray analysis of G272.2$-$3.2. The {\it XMM-Newton} observation was performed with the European Photon Imaging Camera (EPIC), which consists of three detectors, two MOS cameras \citep{turner2001}, and one pn camera \citep{struder2001} operating in the 0.2$-$15~keV range. The satellite was pointed to $\alpha$= 09$^h$06$^m$46$\fs$0 and $\delta$=$-$52$\degr$07$\arcmin$12$\farcs$0 (J2000.0), with the SNR placed at the central CCD. The {\it XMM-Newton} data were analyzed with the {\it XMM-Newton} Science Analysis System (SAS) version 11.0.0. Starting from level-1 event files, the latest calibrations were applied with the ``emproc" and ``epproc" tasks. The events were then filtered to retain only the ``patterns" and photon energies likely for X-ray events: patterns 0 to 4 and energies 0.2 to 15.0 keV for the PN, patterns 0 to 12 and energies 0.2 to 12.0 keV for MOS1/2  instruments. To exclude strong background flares, which could eventually affect the observations, we extracted light curves of photons above 10 keV from the entire field of view of the cameras and excluded intervals up to $3\sigma$ to produce a good time interval (GTI) file.

Two observations from the Advanced CCD Image Spectrometer (ACIS) camera are available in the Chandra archive. ACIS operates in the 0.1$-$10~keV range with high spatial resolution (0.5~arcsec). These observations were calibrated using the CIAO (version 4.1.2) and CALDB (version 3.2.2) packages. Detailed information of the X-ray observations and the instrumental characteristics are given in Table 1.

\section{Results}
\subsection {X-ray images}

The left-hand panel of Fig. 1 shows a composite three-color image of the {\it Chandra} observations. Firstly, the X-ray images were separated into three different energy bands: soft (0.5-1.0 keV) in red, medium (1.0-1.8 keV) in green, and hard (1.8-3.0 keV) in blue. The remnant is undetected above $3.0$~keV. The right-hand panel of Fig. 1 shows a {\it XMM-Newton} X-ray image in the hard energy band (1.8-3.0 keV). The data from MOS1/2 and pn cameras were merged and corrected for exposure and vignetting effects. In our images, north is up and east is to the left. Individual images were binned into 20 pixel arcmin$^{-1}$ and adaptively smoothed by using a Gaussian kernel radius of 3 bins. 

The {\it XMM-Newton} and {\it Chandra} images reveal details of the X-ray morphology of G272.2$-$3.2 that were not noticed in previous X-ray studies (Greiner \& Egger 1993; Greiner, Egger \& Aschenbach 1994; Harrus et al. 2001). The global shape of the SNR is characterized by a circular diffuse X-ray morphology with an angular size of 14 arcmin, but the west-side shows signs of plasma confinement that are probably due to interaction with a denser medium. Embedded within the total diffuse emission, we observe a centrally brightened region with an angular size of 8 arcmin and several X-ray emission enhancements located on its edge. It is interesting to note that on the northern part of the central emission an elongated hard structure seems to exist. This is indicated with an ellipse in Fig.1 (right-hand panel), oriented in the southeast-northwest direction. High surface brightness regions can also be traced on the western part of the remnant. Moreover, the northern X-ray boundary is incomplete and the southeast part of the SNR presents a low X-ray surface brightness. 

\subsection{Mean photon energy map}

\begin{figure}
\includegraphics[width=8.8cm]{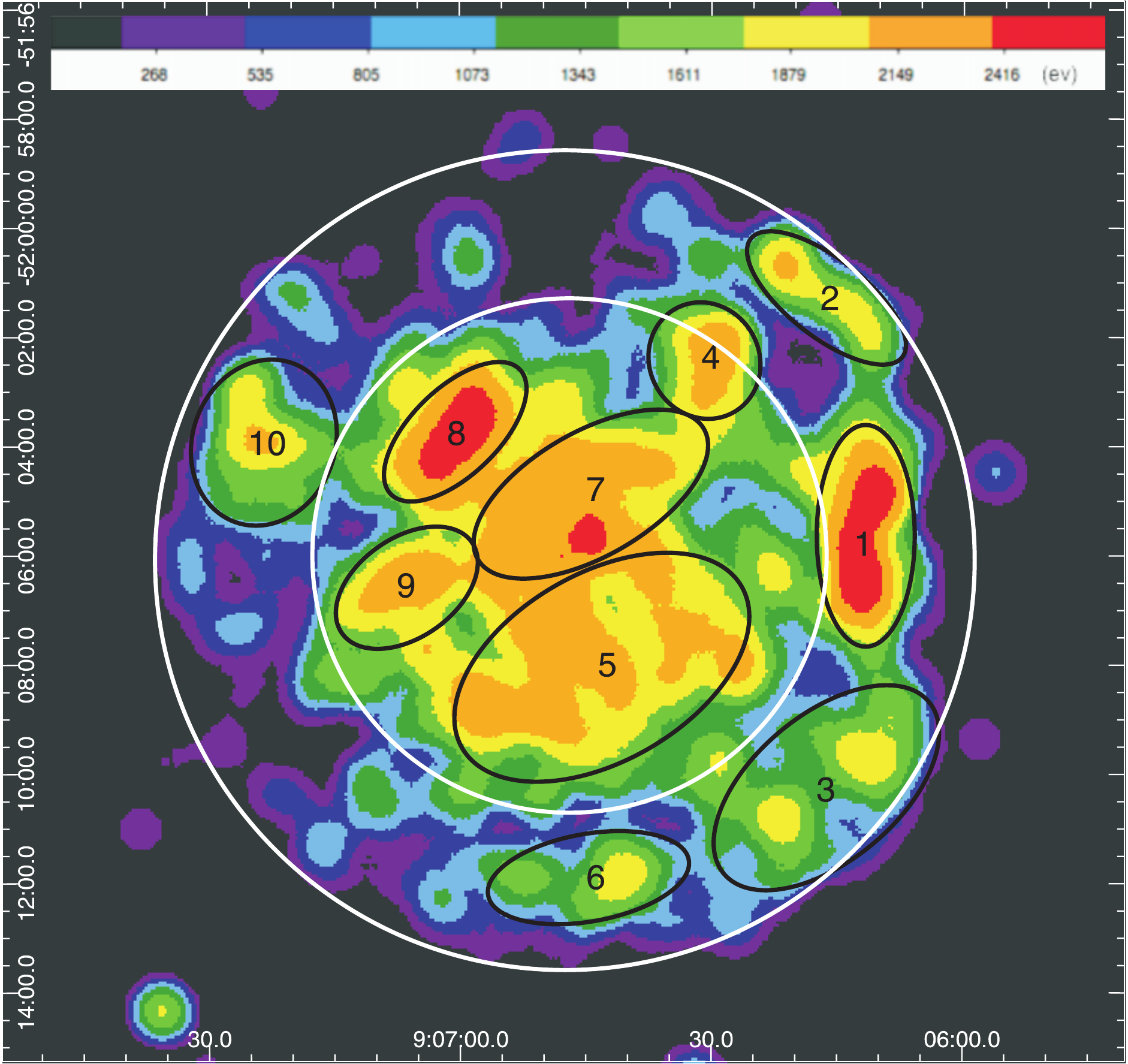}
\caption{MOS mean photon energy map of the 0.3$-$3.0~keV emission (bin size=$9\arcsec$).  Pixels with less than four counts have been masked out. The color bar has a linear scale and the color-coded energy range is between 1.2~keV and 1.7~keV. The X-ray spectra extraction regions selected are indicated in white (for the whole and central part of the SNR) and black circles (for individual regions), respectively.}
\label{mpe}
\end{figure}

To demonstrate the energy dependence of the SNR morphology and to study the possible physical anisotropy of the total X-ray emission, we computed the mean photon energy (MPE) map of the entire SNR (Fig. 2). The MPE map is an image where each computed pixel corresponds to the mean energy of the photons detected by MOS CCDs in the 0.6$-$3.0~keV energy band and thus provides information about the spatial distribution of plasma properties. To compute this map, we merged EPIC MOS1 and MOS2 event files, creating an image with a bin size of $4\arcsec$ that collects a minimum of four counts per pixel everywhere in the remnant. For each pixel, we calculated the mean energy of the photons and then smoothed the map by using a Gaussian kernel value of $3\sigma$ \citep[see][]{miceli2005}.

\subsection{X-ray spectral analysis}

\begin{figure}
\includegraphics[width=8.8cm,angle=0]{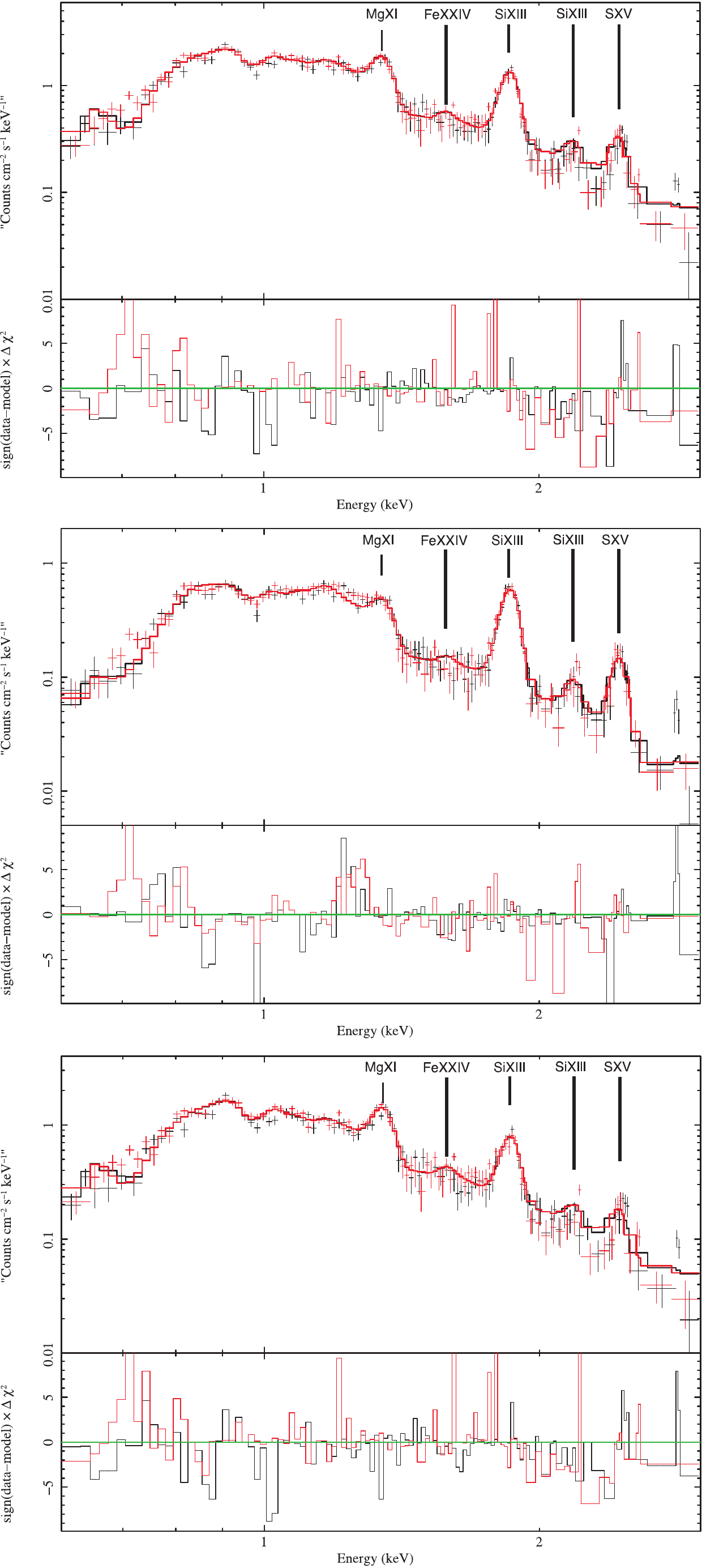}
\caption{{\it XMM-Newton} MOS1/2 spectra of the whole region (upper panel), the central region (central panel), and the outer region (lower panel). Solid lines indicate the best-fit VNEI model (see Table 2). Lower panels present the $\chi^{2}$ fit residuals.}
\label{spectra}
\end{figure}

\begin{table}
\caption{Spectral parameters of the diffuse X-ray emission of selected regions.}
\label{spec}
\renewcommand{\arraystretch}{1.0}
\renewcommand{\tabcolsep}{0.038cm}
\begin{centering}
\begin{tabular}{l| ccc}
\hline\hline
Model \& Parameters & Whole & Center  & Outer \\
\hline
{\bf PHABS*VNEI} &&& \\
N$_\mathrm{H}$ [10$^{22}$~cm$^{-2}$]&	 
0.98$\pm$0.02 &
1.10$\pm$0.01 &
0.87$\pm$0.02  \\
kT [keV] &				 
1.02$\pm$0.06   &
0.76$\pm$0.03   &
1.05$\pm$0.79 
\\
O [O$_\odot$]        &
1.14$\pm$0.19 &
1.0&
0.55$\pm$0.11 
\\
Ne [Ne$_\odot$]  & 
0.50$\pm$0.06 &
0.38$\pm$0.07 &
0.39$\pm$0.05 
\\
Mg [Mg$_\odot$]   &
0.44$\pm$0.04 &
0.44$\pm$0.03 &
0.39$\pm$0.03 
\\
Si [Si$_\odot$] &    
0.84$\pm$0.06 &
1.53$\pm$0.08 &
0.59$\pm$0.05 
\\
S [S$_\odot$] &  
1.60$\pm$0.14&
3.11$\pm$0.24 &
1.05$\pm$0.13 
\\
Fe [Fe$_\odot$]    & 
0.75$\pm$0.06 &
1.34$\pm$0.08 &
0.47$\pm$0.05 
 \\
Ni [Ni$_\odot$]    & 
1.0 &
1.22$\pm$0.75 &
1.0
\\
$\tau$[$10^{10}$~s~cm$^{-3}$] &
3.18$\pm$0.28 &
5.62$\pm$0.80 &
2.84$\pm$0.27 
\\
Norm [$10^{-2}$] & 
3.12$\pm$0.32 &
1.23$\pm$0.80 &
2.33$\pm$0.27 
\\
\hline
$\chi^{2}_{\nu}$ / d.o.f. &
1.70 / 308  &
1.43 / 309 &
1.54 / 308 
\\
\hline
Total Flux(0.6$-$3.0~keV)&
17.18$\pm$0.24&
7.07$\pm$0.29&
9.21$\pm$0.54
\\
\hline
\end{tabular}
\label{spectable}
\tablefoot{Normalization is defined as 10$^{-14}$/4$\pi$D$^2\times \int n_H\,n_e dV$, where $D$ is the distance in [cm], n$_\mathrm{H}$ is the hydrogen density [cm$^{-3}$], $n_e$ is the electron density [cm$^{-3}$], and $V$ is the volume [cm$^{3}$]. Fluxes are absorption-corrected and error values are at the 90\% confidence interval for every single parameter and given in units of 10$^{-11}$~erg~cm$^{-2}$~s$^{-1}$. Abundances are given relative to the solar values of \cite{anders1989}.}
\end{centering}
\end{table}

In order to analyze physical and chemical conditions of the global X-ray emission in G272.2$-$3.2, we extracted spectra from two concentric circular regions with radius 3.8 (central region) and 7.3 (the whole SNR) arcmin, as well as from an annulus region (the outer region), as in Sezer \& G$\ddot{\rm o}$k (2012). In addition, spectra from ten other individual regions, chosen on the basis of the morphology observed in the X-ray images and the MPE map, were analyzed (see Fig. 2). The global (in white) and individual regions (in black) are indicated in Fig. 2. Spectra were obtained using {\sc evselect} SAS task with the appropriate parameters for EPIC MOS 1/2 cameras, and background was subtracted using the {\it XMM-Newton} Blank Sky files \citep{carter2007} for the same regions. For all regions, point-like sources detected using the {\it Chandra} task ``wavdetect" were removed. 

Figure 3 shows the background-subtracted X-ray spectra of the whole, the central, and the outer regions of the SNR. In this figure, the spectra are grouped with a minimum of 16~counts per bin. Error bars are at 90$\%$ confidence level and $\chi^{2}$ statistics is used. The spectral analysis was performed using the XSPEC package \citep{arnaud1996}.

 The whole region (see Fig. 3, upper panel) shows sub-solar abundances of \ion{Ne}{x} (1.27 keV), \ion{Mg}{xi} (1.34 keV), \ion{Si}{xiii} (1.85~keV), and \ion{Fe}{xxiv} with a relatively intense  \ion{S}{xv} (2.4~keV) emission line. The central region (see Fig. 3, central panel) is dominated by weak emission lines of \ion{Ne}{x} and \ion{Mg}{xi}, relatively strong emission lines of \ion{Si}{xiii}, \ion{Fe}{xxiv} and \ion{Ni}{xix} (0.88~keV), and an intense emission line of \ion{S}{xv} (2.4~keV). In contrast, the X-ray spectrum of the outer region displays sub-solar abundances of O, Ne, Mg, Si, and Fe.

The spectra of the regions were fitted with various models: a simple Bremsstrahlung, MEKAL, power-law, NEI, VNEI, PSHOCK, and VPSHOCK, each modified by an absorption interstellar model \citep[PHABS;][]{balucinska1992}. After several tests, the best fit for all regions was computed using a VNEI model, which requires non-solar abundances in some cases for O, Mg, Fe, Si, and S. All other element abundances were fixed at solar values. Our spectral analysis confirms the thermal nature of the emitting plasma found by Greiner \& Egger (1993), Harrus et al. (2001), and Sezer \& G$\ddot{\rm o}$k (2012). The X-ray parameters of the best fit to the diffuse emission spectra for the different regions are presented in Table 2 and Table 3.


The value of the parameters obtained in our fit for the global regions are consistent with those found by Sezer \& G$\ddot{\rm o}$k (2012) using Suzaku observations. For the individual regions, we found that those located on the edge of the remnant (numbered 1, 2, 3, 6, and 10 in Fig. 2) have their X-ray spectra dominated by sub-solar abundances.
In contrast, the X-ray spectra of regions located at the central part of the remnant (numbered 4, 5, 7, 8, and 9 in Fig. 2) are dominated by intense emission lines of  \ion{Si}{xiii}, \ion{S}{xv} (2.4~keV), and \ion{Fe}{xxiv}. It is interesting to note that regions 5 and 9 show large O abundance. This last result is different from the value obtained when fitting the spectrum extracted for the central circle. This suggests that the southern part of the central region has a higher concentration of O than the northern one.

In addition, the spatially resolved spectral analysis shows that the physical conditions of the plasma are not homogeneous throughout the remnant. The values of column density ($N_{\rm H}$) in the regions range approximately from 0.72 $\times$ 10$^{22}$ cm$^{-2}$ to 1.14$\times$ 10$^{22}$ cm$^{-2}$. A lower value was obtained for the outer region, while the highest value of $N_{\rm H}$ corresponds to the central part of the remnant. 

Finally, it is interesting to note that although our fit for the global regions is consistent with those from Sezer \& G$\ddot{\rm o}$k (2012) using Suzaku observations, some differences are noticeable in the abundances. The improved spatial resolution of XMM over Suzaku plays an important role to map small-scale spectral changes in this SNR.  Moreover, Sezer \& G$\ddot{\rm o}$k (2012) used an energy range between 0.3 keV and 10 keV for their analysis, while our study of the XMM-Newton and Chandra data was carried out in a different range, since the remnant is undetected above $3.0$~keV.


\begin{figure*}
\centering
\includegraphics[width=9cm]{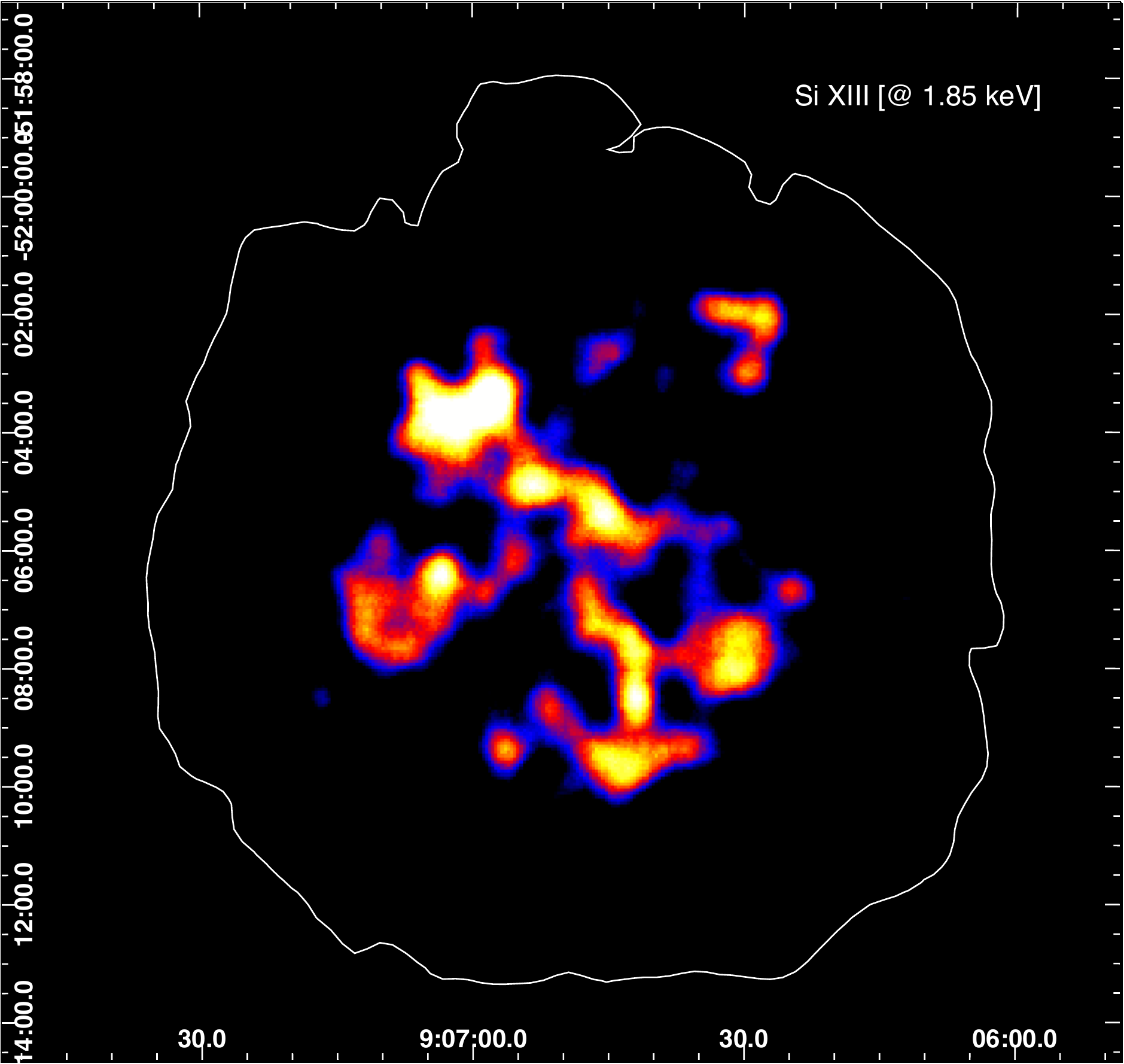}
\includegraphics[width=9cm]{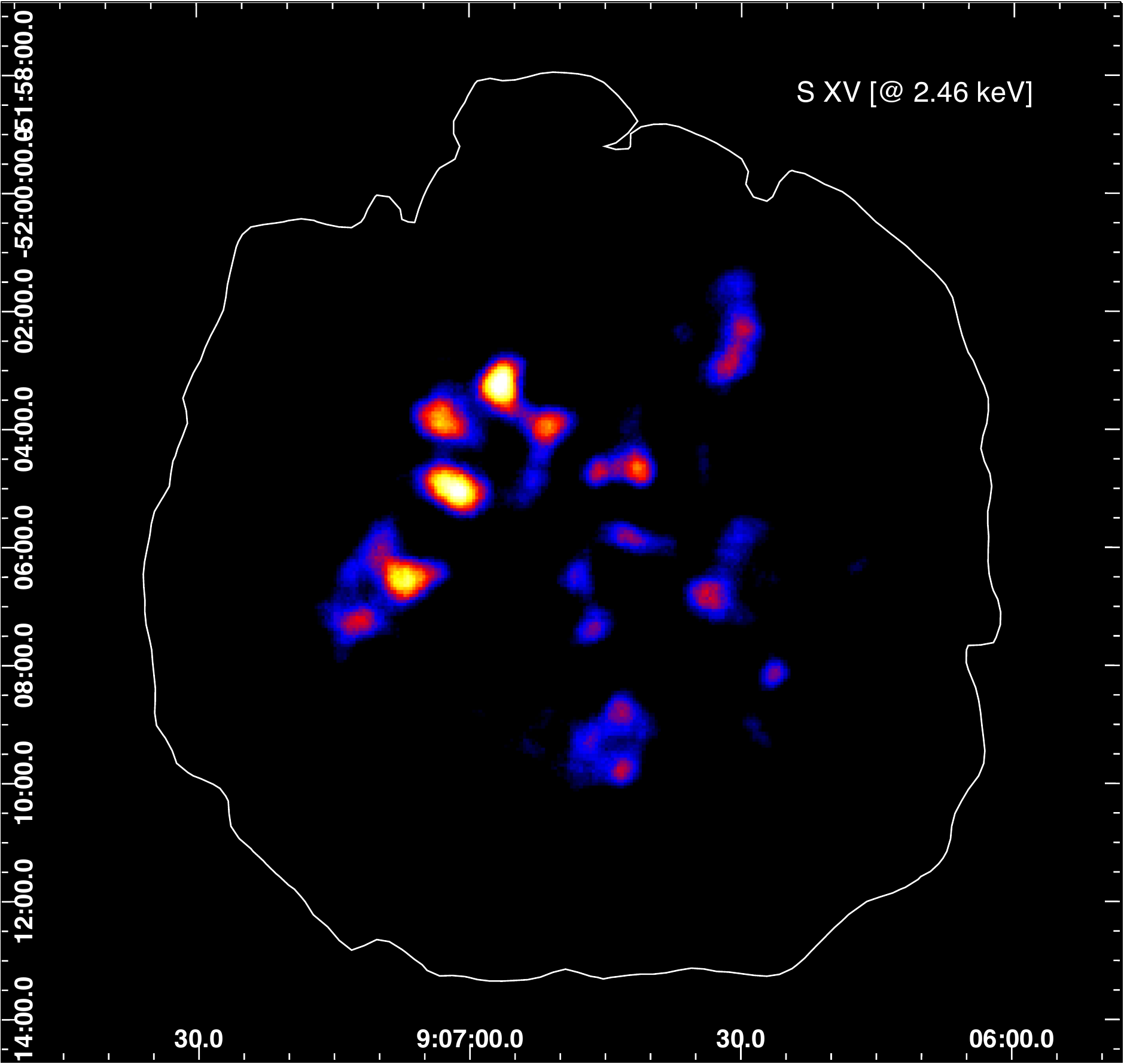}
\caption{Vignetting-corrected {\it XMM-Newton} images of \ion{Si}{xiii} (1.85~keV), and \ion{S}{xv} (2.4~keV). Both images were produced by following the same analysis that was used in Fig 1. Observed X-ray emission from line elements are three sigmas above local background. White contours indicate the peripheral edge of total X-ray emission of G272.2$-$3.2.}
\label{Fig2}
\end{figure*}

\begin{figure*}
\centering
\includegraphics[width=18cm]{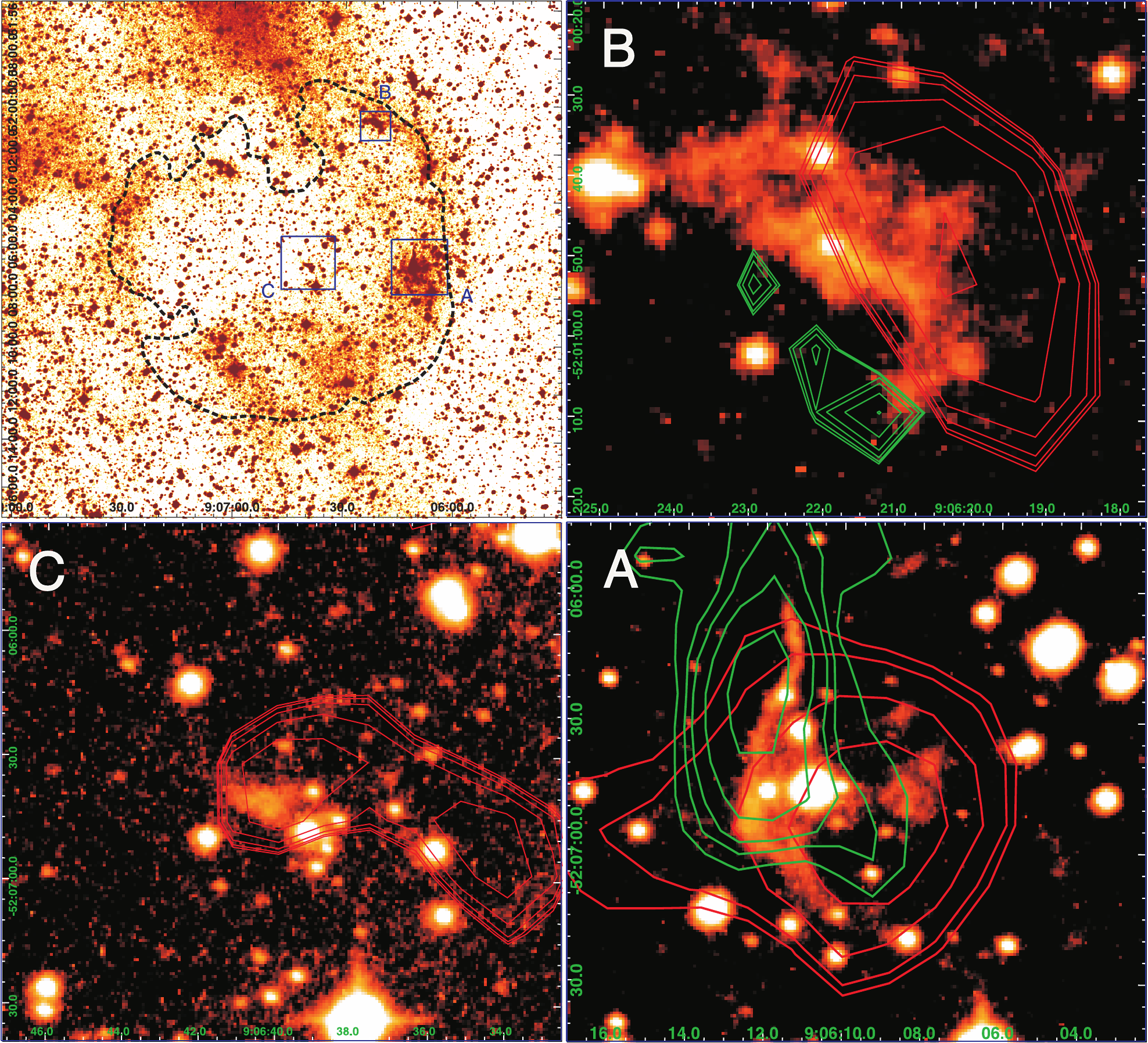}
\caption{SuperCOSMOS H$\alpha$ image of a $25 \times 25$ arcmin$^{2}$ field centered on the center of SNR~G272.2-03.2 in the upper left panel. The outer X-ray contour of the remnant is plotted as a dotted black line. The black boxes mark the location of H$\alpha$ filaments that are plotted in Fig. 5 (A, B, and C). Red and green contours are from radio and X-ray data, respectively.}
\label{halpha}
\end{figure*}

\begin{table*}
\caption{Spectral parameters of the diffuse X-ray emission of the selected regions.}
\label{spec}
\renewcommand{\arraystretch}{1.0}
\renewcommand{\tabcolsep}{0.038cm}
\begin{centering}
\begin{tabular}{l| cccccccccc}
\hline\hline
Model \& Parameters & region 1 & region 2 & region 3 & region 4 & region 5 &  region 6 & region 7  & region 8 & region 9 & region 10\\
\hline
{\bf PHABS*VNEI} &&&&&&&&&& \\
N$_\mathrm{H}$ [10$^{22}$~cm$^{-2}$]&	 
0.77$\pm$0.06 &
0.88$\pm$0.09 &
0.72$\pm$0.05 &
0.97$\pm$0.07 &
1.08$\pm$0.03 &
0.72$\pm$0.07 &
1.08$\pm$0.02 & 
1.07$\pm$0.04 &
1.14$\pm$0.07 &
0.85$\pm$0.07  \\
kT [keV] &				 
0.58$\pm$0.05   &
0.82$\pm$0.16   &
1.19$\pm$0.24   &
0.88$\pm$0.16   &
0.84$\pm$0.07 &
1.41$\pm$0.42   & 
0.66$\pm$0.03 &
0.73$\pm$0.08 &
1.43$\pm$0.41 &
0.85$\pm$0.13 
\\
O [O$_\odot$]        &
0.11$\pm$0.06 &
0.69$\pm$0.42 &
0.42$\pm$0.15 &
0.98$\pm$0.76 &
4.01 $\pm$ 1.95 &
0.28$\pm$0.16 &
1.0 &
1.0 &
4.81$\pm$ 3.03 &
0.34$\pm$0.19 
\\
Ne [Ne$_\odot$]  & 
0.23$\pm$0.06 &
0.72$\pm$0.27 &
0.46$\pm$0.11 &
0.48$\pm$0.26 &
0.99$\pm$0.39 &
0.34$\pm$0.13 &
0.17$\pm$0.12 &
1.0 &
1.02$\pm$0.58 &
0.32$\pm$0.12 
\\
Mg [Mg$_\odot$]   &
0.41$\pm$0.06 &
0.41$\pm$0.11 &
0.58$\pm$0.12 &
0.53$\pm$0.17 &
0.91$\pm$0.27 &
0.49$\pm$0.13 &
0.42$\pm$0.06 &
0.63$\pm$0.14 &
0.80$\pm$0.28 &
0.51$\pm$0.10
\\
Si [Si$_\odot$] &    
0.45$\pm$0.07 &
0.31$\pm$0.10 &
0.65$\pm$0.15 &
1.33$\pm$0.34 &
2.92$\pm$0.82 &
0.48$\pm$0.15 &
1.47$\pm$0.16 &
3.49$\pm$0.63 &
3.04$\pm$0.90 &
0.50$\pm$0.10 
\\
S [S$_\odot$] &  
1.0&
1.05$\pm$0.46 &
0.68$\pm$0.29 &
3.52$\pm$0.89 &
4.79$\pm$1.34 &
0.72$\pm$0.33 &
2.76$\pm$0.39 &
6.36$\pm$1.25 &
4.54$\pm$1.44 &
0.62$\pm$0.31 
\\
Fe [Fe$_\odot$]    & 
0.3$\pm$0.06 &
0.44$\pm$0.14 &
0.58$\pm$0.12 &
0.90$\pm$0.33 &
2.51$\pm$0.81 &
0.45$\pm$0.14 &
1.56$\pm$0.16 &
2.26$\pm$0.48 &
2.80$\pm$0.98 &
0.44$\pm$0.11 
 \\
$\tau$[$10^{10}$~s~cm$^{-3}$] &
5.14$\pm$1.59 &
3.12$\pm$1.08 &
2.03$\pm$0.44 &
5.06$\pm$1.89 &
4.77$\pm$0.82 &
2.07$\pm$0.47 &
8.00$\pm$2.46 &
6.70$\pm$2.20 &
2,30$\pm$0.42 &
3.14$\pm$0.82 
\\
Norm [$10^{-3}$] & 
4.60$\pm$1.60 &
1.44$\pm$0.98 &
1.55$\pm$0.79 &
0.83$\pm$0.27 &
2.15$\pm$0.52 &
0.69$\pm$0.24 &
3.02$\pm$0.28 &
0.80$\pm$0.10 &
0.37$\pm$0.26 &
1.76$\pm$0.53 
\\
\hline
$\chi^{2}_{\nu}$ / d.o.f. &
1.08 / 225  &
0.95 / 211 &
1.09 / 261 &
0.97 / 216 &
1.27 / 296 &
1.03 / 229 &
1.04 / 273 &
1.30 / 235 &
1.08 / 239 &
1.01 / 233

\\
\hline
Flux(0.6$-$1.0~keV)& 
42.50$\pm$0.72&
66.96$\pm$1.58&
34.28$\pm$0.51&
25.79$\pm$0.59&
194.13$\pm$1.95&
16.25$\pm$0.34&
69.98$\pm$0.96&
34.82$\pm$0.68&
82.72$\pm$1.67&
32.38$\pm$0.64\\
Flux(1.0$-$1.8~keV) & 
9.82$\pm$0.16&
8.24$\pm$0.19&
9.65$\pm$0.14&
5.16$\pm$0.12&
25.69$\pm$0.26&
4.57$\pm$0.09 &
14.70$\pm$0.20 &
7.02$\pm$0.14 &
8.13$\pm$0.16&
7.26$\pm$0.14\\
Flux(1.8$-$3.0~keV) & 
1.64$\pm$0.02&
1.13$\pm$0.02&
2.53$\pm$0.04&
1.86$\pm$0.04&
7.58$\pm$0.07&
1.30$\pm$0.03&
3.81$\pm$0.05&
2.44$\pm$0.05&
2.38$\pm$0.05&
1.62$\pm$0.03\\
\hline
Total Flux(0.6$-$3.0~keV)&
53.50$\pm$0.90&
76.33$\pm$1.39&
46.47$\pm$0.69&
32.81$\pm$0.75&
227.40$\pm$2.28&
22.12$\pm$0.46&
88.49$\pm$1.21&
44.28$\pm$0.87&
93.24$\pm$1.9&
41.26$\pm$0.81\\
\hline
\end{tabular}
\label{spectable}
\tablefoot{Normalization is defined as 10$^{-14}$/4$\pi$D$^2\times \int n_H\,n_e dV$, where $D$ is distance in [cm], n$_\mathrm{H}$ is the hydrogen density [cm$^{-3}$], $n_e$ is the electron density [cm$^{-3}$], and $V$ is the volume [cm$^{3}$]. Fluxes are absorption-corrected, while error values are in the 90\% confidence interval for every single parameter and are given in units of 10$^{-13}$~erg~cm$^{-2}$~s$^{-1}$. Abundances are given relative to the solar values of \cite{anders1989}.}
\end{centering}
\end{table*}

The spatial variation of the Si and S between the central and outer regions is better observed in Fig. 4. These images reveal a non-uniformity in the spatial distribution of the elements, which are mostly concentrated in the central circular region.

\subsection {Spatial correlation between optical and X-ray emission}


When optical emission from SNRs is not obscured by intervening dust, especially at low Galactic latitude, sensitive H$\alpha$ observations can reveal optical filaments or emission structures that are highly correlated with prominent radio or X-ray emission. In order to study the spatial correlation of optical and X-ray observations of G272.2$-$3.2, we extracted an image from the SuperCOSMOS H$\alpha$ survey\footnote{http://www-wfau.roe.ac.uk/sss/halpha/} (Parker et al. 2005). Figure 5 shows the H$\alpha$ image located at the center of the supernova remnant. The external contour of the SNR X-ray emission is overplotted as a black dotted line. Radio contours at 843 MHz (Whiteoak \& Green 1996) were also overplotted in red in Figure 5, A, B, and C. The image was processed to increase contrast. The H$\alpha$ emission is mostly concentrated at the edge of the SNR, showing a cavity at the middle that is coincident with the central X-ray emission. We also present a zoom to each H$\alpha$ filament, overplotted with radio (in red) and X-ray contours (in green). The filaments indicated as A and B coincide partially with radio and X-ray emission peaks.

In both images, the peak of the radio emission lies to the inner side of the H$\alpha$ emission and the peak of the X-ray emission to the outer side. The H$\alpha$ morphology of filament A (in particular the concavity pointing outward) is reminiscent of the one in the Vela SNR D filament observed by Bocchino et al. (2000), while filament B is more similar to filament E of the same work. The Vela SNR filaments D and E of Bocchino et al. (2000) have been interpreted in terms of shocks encountering a small isolated cloud and a larger feature, respectively. Similar to what happens in the case of the Vela filaments D and E, our filaments A and B also show X-ray emission peaks inside the optical filament, i.e., the X-ray peak is nearer to the center of the remnant. This is in agreement with the scenario of the shock encountering a strong density gradient along the expected direction of propagation (east to west), as argued by Bocchino et al. (2000) for the Vela filaments.

Although the results suggest that the H$\alpha$ emission could be associated with part of the remnant, this interpretation should be taken with caution, since it may be associated with foreground nebulosity located to the north and north-east and therefore physically unrelated to the remnant. A more detailed study of the filaments is needed to address this issue.

\section {Discussion}

The spatially resolved spectral analysis MPE map and available information obtained at radio and H$\alpha$ frequencies of G272.2$-$3.2 permit a realistic astrophysical scenario to be outline, allowing us to better understand the remnant's evolution and the observed properties at all wavelengths.

The X-ray morphology of the remnant displays several interesting characteristics. It comprises a global structure of almost circular X-ray emission and a centrally brightened hard region with a peculiar elongated structure in the upper part that is orientated in the northwest-southeast direction. In Fig. 1 (right-hand panel), these components were already pointed out. The X-ray spectral study of the regions shows that there are distinct K$\alpha$ emission-line features of metal elements, confirming the thermal origin of the emission. Moreover, the absorbed VNEI thermal plasma model used (see Table 2) produces elevated abundances of Si, S and Fe in the circular central region, typical of ejecta material. In contrast, the outer region shows under-solar abundances, which are to be expected if the emission arises from the shocked ISM. The relatively low ionization timescale ($\leq$ 10$^{11}$ cm$^{-3}$ s) suggests that the plasma is not in ionization equilibrium.

In order to analyze the behavior of the physical parameters obtained from the spatially resolved spectral study, we performed a series of plots of these parameters versus the distance from the center of the SNR. Figure 6 shows the ionization timescale, temperature, electron density, neutral hydrogen absorption column, and individual abundances as a function of the distance from the geometrical center of the SNR outwards. 

High values of $\tau$ correspond to regions 7 and 8, which are regions belonging to the central part of the SNR, where an elongated hard X-ray feature is observed. The temperatures of the plasma range from 0.5 to 1.4 keV, with the highest values located in regions 6 and 9. The electronic density (computed for the range of distance between 5-10 kpc) is high in regions 1, 2, and 6 on the northwest (where filamentary H$\alpha$ emission is observed, i.e., regions A and B in Figure 5) and southern parts of the SNR. This is in agreement with the idea that the remnant is encountering more dense material at these locations. Abundances of O, Si, S, and Fe are also shown in Fig. 6, lower panel. These plots allow us to infer the spatial distribution of the stellar ejecta in the SNR. It is interesting to note that O is high in regions 5 and 9, and Si, S, and Fe are high in regions 4, 5, 7, 8, and 9. All these regions lie inside the central part of the SNR. These results confirm that the X-ray emission inside the central region of the SNR is associated with  material of the ejecta. The average value of Mg/Si, S/Si, and Fe/Si in the inner regions is approximately 0.28, 1.89, and 0.83, respectively. The low O and high Fe abundances suggests a Type Ia progenitor, which is also in rough agreement with the expected S/Si and Fe/Si value from detonation models (Badenes et al 2003). Very little Mg is expected for Type Ia SN, which primarily produces Fe-group elements, and the absolute Mg abundances are indeed low in the central region, even if the Mg/Si is higher than expected (but with high uncertainties). So the overall picture that is favored is an interpretation in terms of a Type Ia progenitor, even if the high O abundances in regions 5 and 9 cast some doubts on this. 

\begin{figure*}
\includegraphics[width=18cm]{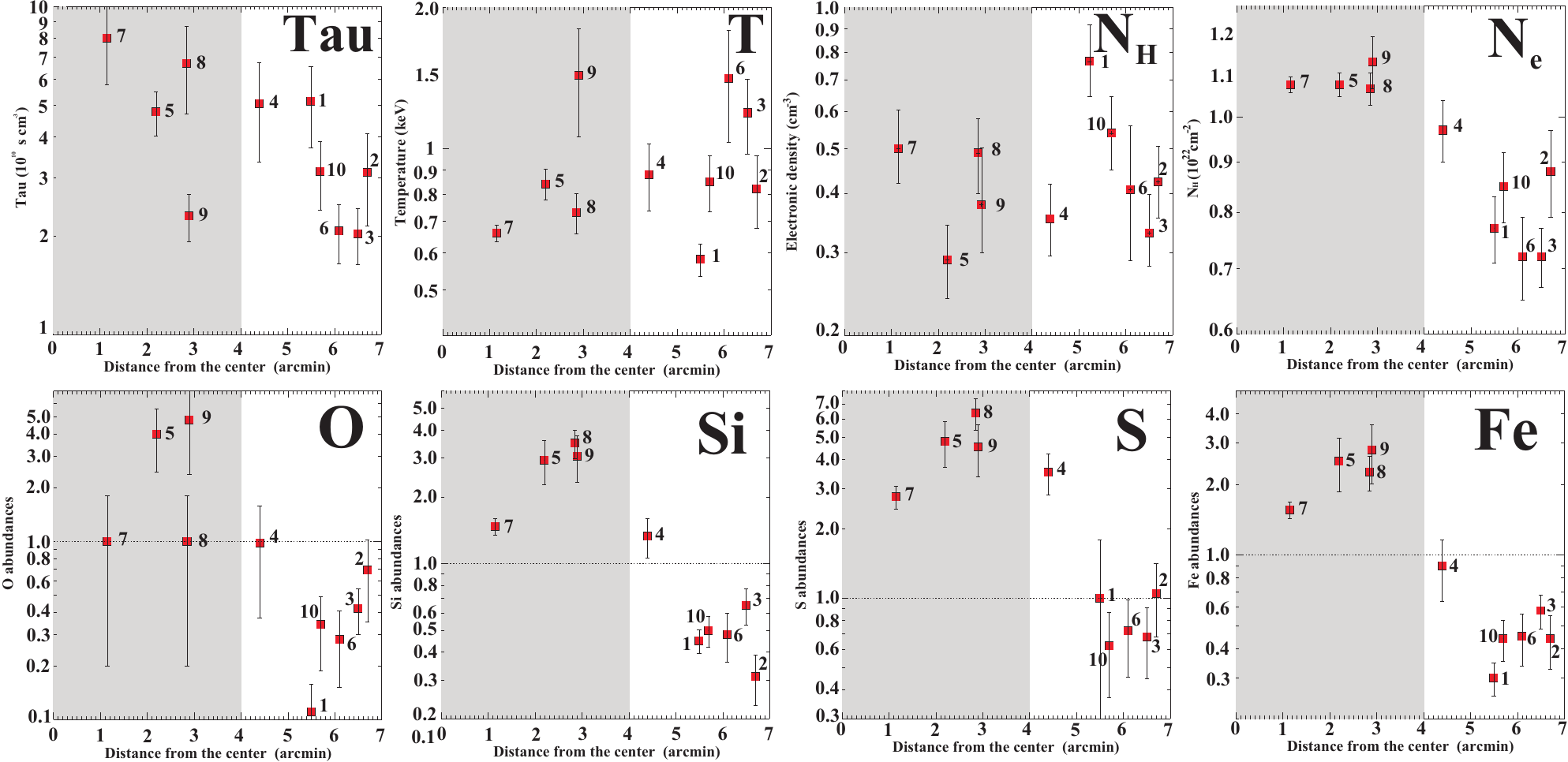}
\caption{X-ray parameters obtained from the spectral analysis as a function of the distance from the geometric center of the SNR. Error bars in the electronic density represent the distance in the range of 5-10 kpc.  The region of the reverse shock is highlighted in grey. In the lower plots, the dotted line indicates the solar abundance.}
\label{Fig2}
\end{figure*}

With the aim to assess the relative abundances in the ejecta, we performed a plot of the abundances relative to Si. Figure 7 shows a comparison of our best-fitting relative abundances with the Suzaku results, the nucleosynthesis predicted by the widely used Nomoto et al. (1997) model, and a delayed detonation Type Ia SN model (Nomoto et al. 1997).  For the abundance ratio of Ca/Si, we used the value obtained by Sezer \& G$\ddot{\rm o}$k (2012) with Suzaku observations, since the Ca emission lines cannot be measured in our spectra and therefore its abundance cannot be determined. Our fit is mostly consistent with the Suzaku data and both models, although the abundance of Ni relative to Si is lower than the value obtained by Sezer \& G$\ddot{\rm o}$k (2012) and higher than what both models predict. These results confirm that G272.2$-$3.2 has a Type Ia SN origin.

The results also suggest that both the outer and inner regions represent shock waves propagating outward (the forward shock is characterized by the radius R$_{\rm S}$ in Figure 1, right-hand panel), and inward (the reverse shock), respectively. The interface between the two regions is the contact discontinuity (at R$_{\rm CD}$ in Figure 1). Therefore, while the outermost regions with R$\geq$R$_{\rm CD}$ are likely composed of shocked interstellar material (and a small ejecta component, maybe due to Rayleigh-Taylor instabilities at the contact discontinuity (CD)), the innermost region (with a R $\leq$ R$_{\rm CD}$) is composed of shocked ejecta material.

Using a distance of 5 kpc as a scaling factor for G272.2$-$3.2, the radii of the forward shock R$_{\rm S}$ and the contact discontinuity R$_{\rm CD}$ correspond to 10.9 D$_{\rm 5 kpc}$ pc and 5.8 D$_{\rm 5 kpc}$ pc, respectively. If the remnant expands in a uniform ambient medium or a medium with a power-law density profile with index $-$2 with an ejecta profile also in the form of a steep power-law, Chevalier (1982) has demonstrated that the ratio  $\psi = $R$_{\rm S}$/R$_{\rm CD}$ cannot exceed the value 1.3-1.4 under all reasonably realistic circumstances. This is confirmed by a recent numerical work on this subject, which also shows that Rayleigh-Taylor instabilities at the CD cause ejecta material to protrude into the ISM-dominated post-shock zone, thus getting even closer to the forward shock (Orlando et al. 2012). It is therefore not straightforward to interpret our observed value of the ratio $\psi$ ($\sim 1.8-1.9$) in terms of classical models. One possibility could be that the expansion occurred in a more dense medium during the early evolutionary phase and the forward shock only recently encountered a less dense medium, thus accelerating to higher velocity and increasing its distance from the shocked ejecta still in the more dense part. This situation is similar to remnants expanding in the dense Red Supergiant (RSG) wind that surrounds the core-collapse supernova (SN). We point out that this may not be at odds with the Type Ia origin proposed above because progenitors in the single degenerate scenarios are expected to have fast optical thick outflows from the white-dwarf surface (Hachisu et al. 2012). By making the simple and crude assumption that the ejecta have just arrived at the interface between the two media and by assuming a constant $\rho v^2$, we may infer that the difference between the expected and observed value of $\psi$ may be reproduced by a factor 2 density jump at the interface between the two media. Unfortunately, the X-ray spectral fits do not give significant density variations between the outer and inner regions to be fully consistent with such a hypothesis. However, we do find an apparent increase by a factor 2 for the interstellar absorption in the central regions, which may indicate denser foreground material. Such a scenario, if confirmed, could have interesting consequences in the study of the overionization in SNRs. In fact, it was recently claimed that this scenario is responsible for the rapid cooling and overionization conditions observed in some SNRs (Shimizu et al. 2012), while another group has suggested that the dense environment and low density in the interior are the key to understanding the intriguing phenomenon of plasma overionization (Zhou et al. 2011). We have found no signs of recombination edges in our X-ray spectra of G272.

Based on the Sedov (1959) model and using the effective temperature ($T_{\rm eff} \sim$ 1.02 keV) obtained from the spectral analysis, the shock front velocity $V_{\rm sh}$ and the total swept mass $M_{\rm tot}$ can be obtained from the equations derived by Bocchino \& Bandiera (2003):

\begin{equation}
V_{\rm sh}^{2} = k T_{\rm eff} / 0.14 m_{\rm H}
\end{equation}

\noindent and

\begin{equation}
M_{\rm tot} = 4.19 \rho_{\rm 0} R_{\rm sh}^{3},
\end{equation}

where \noindent $R_{\rm sh}$ is R$_{\rm S}$ = 10.9 D$_{\rm 5 kpc}$ pc and $\rho_{\rm 0}$ = 1.26 $m_{\rm H}$$\cdot$$n_{\rm 0}$. Here, $m_{\rm H}$ is the hydrogen-atom mass and $n_{\rm 0}$ the density of the ambient medium, that we assumed to be $n_{\rm 0}$= $n_{\rm e}$/4 = 0.1 cm$^{-3}$. The electronic density of the plasma $n_{\rm e}$ was obtained from the X-ray image. Assuming that the plasma fills a region like the external circle indicated in Fig. 2 for a range of distance of 5-10 kpc, we obtain an average volume $\bar{V}$ for the SNR. Thus, based on the emission measure (EM) determined by the spectral fitting (see Table 2), we estimated the electron density of the plasma, $n_{e}$=$\sqrt{EM/V}$$\sim$ 0.46 cm$^{-3}$. In this case, the number density of the nucleons was simply assumed to be the same as that of the electrons. As a result, the shock front velocity is $V_{\rm sh} \sim$  820 km s$^{-1}$ and the total swept mass $M_{\rm tot} \sim$ 17 $M_{\sun}$. 

In this model, the shock has been modeled on the assumption that it is a strong non-radiative shock with a temperature equilibrium between ions and electrons.  However, for a shock with a speed of $\sim$ 800 km s$^{-1}$, equilibrium is unlikely (Ghavamian et al. 2007). Using the electron-to-proton temperature ratio at the shock front as a function of shock velocity for five Balmer-dominated SNRs (see Fig. 2 in Ghavamian et al. 2007), the range of shock velocities permissible, both with and without the equilibrium assumption, is between 800 and 1600 km s$^{-1}$. In this case, the SNR age $t_{\rm snr}$ is

\begin{equation}
t_{\rm snr} = 4.3 D_{\rm 5 kpc} (\rm pc)/ V_{\rm sh} ,
\end{equation}

\noindent that we estimate as  $t_{\rm snr}$  = (2500-5000) D$_{\rm 5 kpc}$ yr.

\begin{figure}
\includegraphics[width=9cm]{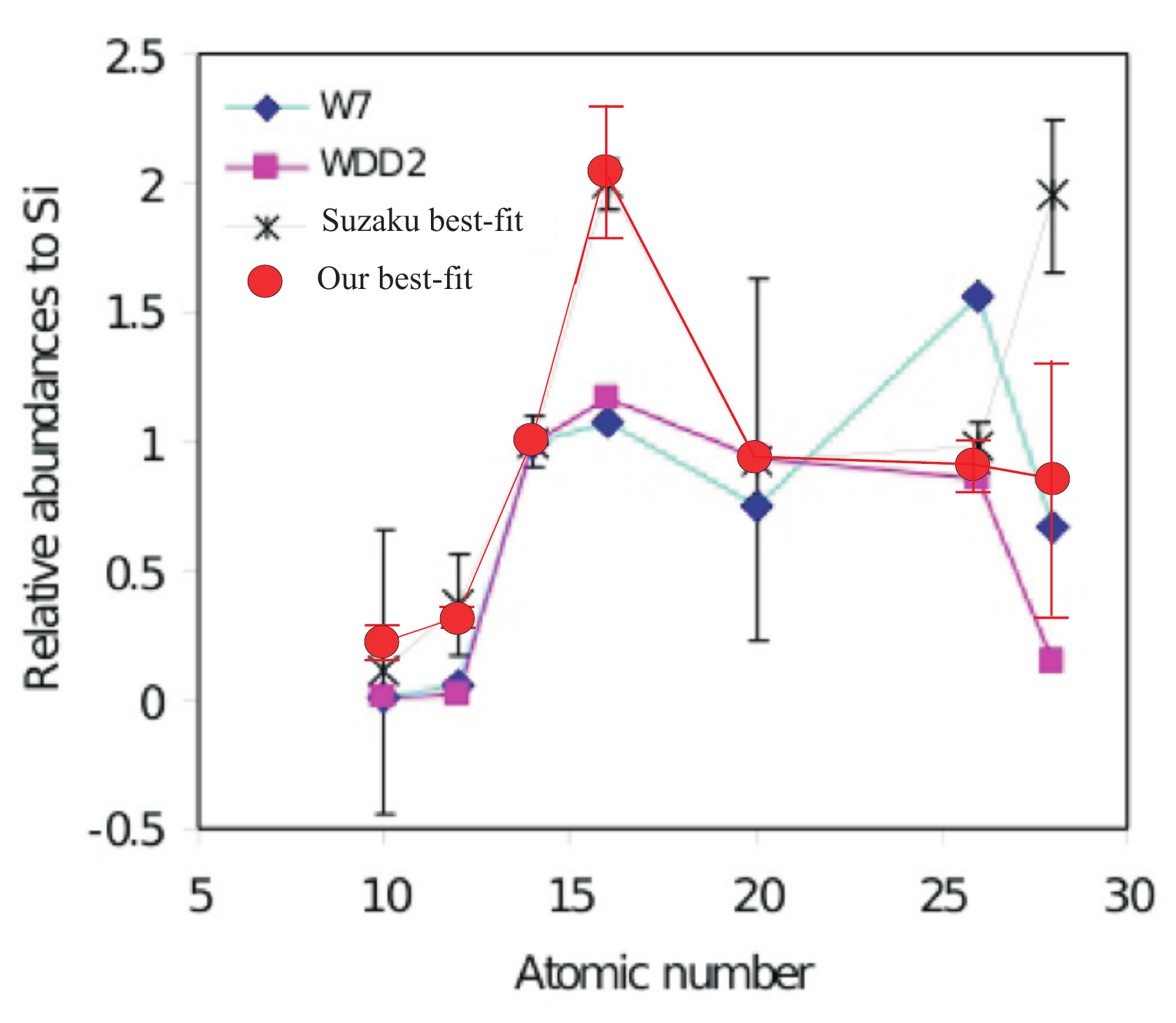}
\caption{The best-fit abundance ratios of Ne, Mg, Si, S, Fe, and Ni relative to Si are shown by red circles. The predicted abundance ratios from the carbon deflagration (W7; Nomoto et al. 1997) model are shown by diamonds and those from the delayed detonation (WDD2; Nomoto et al. 1997) model are shown by squares. The best fit of relative abundances obtained by Sezer \& G$\ddot{\rm o}$k (2012) is also indicated.  For the abundance ratio of Ca/Si, we assumed the value obtained by Sezer \& G$\ddot{\rm o}$k (2012) using Suzaku observations, since the Ca emission lines cannot be measured in our spectra and therefore its abundance cannot be determined.}
\label{Fig2}
\end{figure}

On the other hand, we can compute the age of the SNR using the information obtained from the spectral analysis. For an average distance between 5-10 kpc  and an ionization timescale $\bar{\tau}$= 3.2$\times$10$^{10}$ cm$^{-3}$ s, we determine that the elapsed time $t$ after the plasma was heated is $t$ $\sim$ 2000 yr.  This value is compatible with a distance of 2.5 kpc (i.e., D$_{\rm 5 kpc}$ = 0.5)  and a velocity of 800 km s$^{-1}$ for the shock front, or a distance of 5 kpc (i.e., D$_{\rm 5 kpc}$ = 1) and a velocity of 1600 km s$^{-1}$ using Sedov's equations.
Therefore, we suggest that the SNR's age likely lies in the range of $t_{\rm snr}$ $\sim$ 3600 ($\pm$ 1600) yr.

\section{Conclusions}

We have used {\it XMM-Newton} and {\it Chandra} observations of G272.2$-$3.2 to study the characteristics of its X-ray emission and the physical connexion with radio and H$\alpha$ observations of the object. The X-ray morphology reveals an overall structure of almost circular X-ray emission, a centrally brightened hard region, and a peculiar elongated hard structure. The X-ray spectral analysis of the regions show emission-line features of metal elements, confirming the thermal origin of the emission. The X-ray spectra are well represented by an absorbed VNEI thermal plasma model, which produces elevated abundances of Si, S, and Fe in the circular central region, typical of ejecta material. The values of abundances found in the central region of the SNR favor a Type Ia progenitor for this remnant, in agreement with the results obtained by Sezer \& G$\ddot{\rm o}$k (2012), using Suzaku observations. In contrast, the outer region shows abundances below the solar value, to be expected if the emission arises from the shocked ISM. We have identified the location of the contact discontinuity and its distance from the outer shock is higher than expected for expansion in a uniform media, suggesting that the remnant spent most of its time in a more dense medium. 



\begin{acknowledgements}
      
We are grateful to the referee for her/his valuable suggestions and comments, which helped us to improve the paper. 
The authors acknowledge support by DGI of the Spanish Ministerio de Educaci\'on y Ciencia under grants AYA2010-21782-C03-03, FEDER funds, Plan Andaluz de Investigaci\'on Desarrollo e Innovaci\'on (PAIDI) of Junta de Andaluc\'{\i}a as research group FQM-322, and the excellence fund FQM-5418. J.A.C. and J.F.A.C. are researchers of CONICET. J.F.A.C was supported by grant PIP 2011-0100285 (CONICET). J.A.C by grant PICT 2008-0627, from ANPCyT and PIP 2010-0078 (CONICET). J.L.S. acknowledges support by the Spanish Ministerio de Innovaci\'on y Tecnolog\'ia under grant AYA2008-06423-C03-03. F.B. acknowledges partial support of the ASI-INAF agreement n. I/009/10/0.

\end{acknowledgements}


\begin{thebibliography}{}

\bibitem[Anders \& Grevesse(1989)]{anders1989} Anders, E., \& Grevesse, N.\ 1989, \gca, 53, 197 

\bibitem[Arnaud(1996)]{arnaud1996} Arnaud, K.~A.\ 1996, Astronomical Data Analysis Software and Systems V, 101, 17

\bibitem[Badenes et al.(2003)]{2003ApJ...593..358B} Badenes, C., Bravo, E., Borkowski, K.~J., \& Dom{\'{\i}}nguez, I.\ 2003, \apj, 593, 358

\bibitem[Balucinska-Church \& McCammon(1992)]{balucinska1992} Balucinska-Church, M., \& McCammon, D.\ 1992, \apj, 400, 699 

\bibitem[Bocchino et al.(2000)]{2000A&A...359..316B} Bocchino, F., Maggio, A., Sciortino, S., \& Raymond, J.\ 2000, \aap, 359, 316 

\bibitem[Bocchino \& Bandiera(2003)]{2003A&A...398..195B} Bocchino, F., \& Bandiera, R.\ 2003, \aap, 398, 195 

\bibitem[Boumis et al.(2002)]{2002A&A...385.1042B} Boumis, P., Mavromatakis, F., \& Paleologou, E.~V.\ 2002, \aap, 385, 1042

\bibitem[Carter \& Read(2007)]{carter2007} Carter, J.~A., \& Read, A.~M.\ 2007, \aap, 464, 1155

\bibitem[Chevalier(1982)]{1982ApJ...258..790C} Chevalier, R.~A.\ 1982, \apj, 258, 790

\bibitem[Duncan et al.(1997)]{1997MNRAS.287..722D} Duncan, A.~R., Stewart, R.~T., Haynes, R.~F., \& Jones, K.~L.\ 1997, \mnras, 287, 722 

\bibitem[Ghavamian et al.(2007)]{2007ApJ...654L..69G} Ghavamian, P., Laming, J.~M., \& Rakowski, C.~E.\ 2007, \apjl, 654, L69 

\bibitem[Green(2009)]{2009BASI...37...45G} Green, D.~A.\ 2009, Bulletin of the Astronomical Society of India, 37, 45 

\bibitem[Greiner \& Egger(1993)]{1993IAUC.5709....2G} Greiner, J., \& Egger, R.\ 1993, \iaucirc, 5709, 2 

\bibitem[Greiner et al.(1994)]{1994A&A...286L..35G} Greiner, J., Egger, R., \& Aschenbach, B.\ 1994, \aap, 286, L35

\bibitem[Hachisu et al.(2012)]{2012ApJ...744...69H} Hachisu, I., Kato, M., Saio, H., \& Nomoto, K.\ 2012, \apj, 744, 69

\bibitem[Harrus et al.(2001)]{2001ApJ...552..614H} Harrus, I.~M., Slane, P.~O., Smith, R.~K., \& Hughes, J.~P.\ 2001, \apj, 552, 614 

\bibitem[Koo \& Kang(2004)]{2004MNRAS.349..983K} Koo, B.-C., \& Kang, J.-h.\ 2004, \mnras, 349, 983 

\bibitem[Mavromatakis \& Strom(2002)]{2002A&A...382..291M} Mavromatakis, F., \& Strom, R.~G.\ 2002, \aap, 382, 291 

\bibitem[Mavromatakis et al.(2005)]{2005A&A...435..141M} Mavromatakis, F., Boumis, P., Xilouris, E., Papamastorakis, J., \& Alikakos, J.\ 2005, \aap, 435, 141

\bibitem[Miceli et al.(2005)]{miceli2005} Miceli, M., Bocchino, F., Maggio, A., \& Reale, F.\ 2005, \aap, 442, 513 

\bibitem[Nomoto et al.(1997)]{1997NuPhA.621..467N} Nomoto, K., Iwamoto, K., 
Nakasato, N., et al.\ 1997, Nuclear Physics A, 621, 467

\bibitem[Orlando et al.(2012)]{2012ApJ...749..156O} Orlando, S., Bocchino, 
F., Miceli, M., Petruk, O., \& Pumo, M.~L.\ 2012, \apj, 749, 156

\bibitem[Parker et al.(2005)]{2005MNRAS.362..689P} Parker, Q.~A., Phillipps, S., Pierce, M.~J., et al.\ 2005, \mnras, 362, 689

\bibitem[Sedov(1959)]{1959sdmm.book.....S} Sedov, L.~I.\ 1959, Similarity and Dimensional Methods in Mechanics, New York: Academic Press, 1959


\bibitem[Sezer \& G$\ddot{\rm o}$k(2012)]{2012MNRAS.421.3538S} Sezer, A., \& G$\ddot{\rm o}$k, F.\ 2012, \mnras, 421, 3538

\bibitem[Shimizu et al.(2012)]{2012PASJ...64...24S} Shimizu, T., Masai, K., \& Koyama, K.\ 2012, \pasj, 64, 24

\bibitem[Str{\"u}der et al.(2001)]{struder2001} Str{\"u}der, L., et al.\ 2001, \aap, 365, L18

\bibitem[Turner et al.(2001)]{turner2001} Turner, M.~J.~L., et al.\ 2001, \aap, 365, L27 

\bibitem[Whiteoak \& Green(1996)]{whiteoak1996} Whiteoak, J.~B.~Z., \& Green, A.~J.\ 1996, \aaps, 118, 329 

\bibitem[Zhou et al.(2011)]{2011MNRAS.415..244Z} Zhou, X., Miceli, M., Bocchino, F., Orlando, S., \& Chen, Y.\ 2011, \mnras, 415, 244 

\end{thebibliography}
\end{document}